Title: Off-center impurity in alkali halides: reorientation, electric polarization and pairing to F center. IV. Reorientational rate.
Authors: G. Baldacchini (1), R.M. Montereali (1), U.M. Grassano (2), A. Scacco (3), P. Petrova (4), M. Mladenova (5), M. Ivanovich (6), and M. Georgiev (6) (ENEA C.R.E. Frascati, Frascati (Roma), Italy (1), Dipartimento di Fisica, Universita "Tor Vergata", Rome, Italy (2), Dipartimento di Fisica, Universita "La Sapienza", Rome, Italy (3), Institute of Nuclear Research and Nuclear Energy, Bulgarian Academy of Sciences, Sofia, Bulgaria (4), Department of Condensed Matter Physics, Faculty of Physics, University of Sofia, Sofia, Bulgaria (5), Institute of Solid State Physics, Bulgarian Academy of Sciences, Sofia, Bulgaria (6))
Comments: 17 pages including 2 figures and 4 tables, all pdf format
Subj-class: physics

This last Part IV is aimed at deriving relaxation rates (times) of an off-centered $Li^+$ impurity. We follow Christov's reaction-rate method to define general rate equations in terms of the exact Mathieu eigenvalues, as well as of harmonic-oscillator eigenvalues approximating for the energy spectrum near the bottom of the reorientational wells. To calculate the rate in each particular case we derive configurational tunneling probabilities by either Mathieu functions or by harmonic-oscillator functions. The electron-transfer probability is calculated by generalizing Landau-Zener's method. Typical examples are considered and compared with experimental relaxation times in alkali halides.

1. Introduction

Following the preceding Part I through III [1-3], this Part IV deals with the relaxation time, inversely proportional to the relaxation rate of an off-centered impurity. The problem arises as the system relaxes to thermal equilibrium following an initial disturbance which overpopulates some reorientational sites at the expense of others [4]. In as much as an off-centered impurity and its immediate surroundings can be regarded as a strongly-coupled electron-vibrational mode system, we calculate a relaxation rate by the reaction-rate method [5]. For this purpose we redefine the general two-site rate equation for two cases of in-plane reorientation of the $Li^+$ impurity by using (i) the exact Mathieu eigenvalues and (ii) the harmonic-oscillator eigenvalues approximating for the energy spectrum near the bottom of each reorientational well. We then calculate configurational tunneling probabilities using eigenfunctions corresponding to the exact and approximate eigenvalues. While tunneling reorientations in the harmonic approximation have been described previously [6], a Mathieu-based analysis is now reported for the first time. In both cases, the electron-transfer probability is calculated following Christov's generalization of Landau-Zener's method [5].

Experimental data on the relaxation times of the $Li^+$ impurity are presently available for the isolated case only [7]. In spite of that we also present calculated examples of theoretical rates with the hope of stimulating further experiments.

## 2. Reorientational rate

### 2.1. Reorientation of off-center impurity

Factorizing the wavefunction $\Phi(Q_l) = \Pi_i \Phi_i(Q_i)$ using $E = \Sigma_i E_i$, Schrödinger's equation for motion along a reorientational ring

$$[-(\eta^2/2M)\Sigma_i(\partial^2/\partial Q_i^2)+(d_b-d_c)(M\omega_n^2/b)\Sigma_i Q_i^4]\Phi(Q_l)= E\, \Phi(Q_l) \qquad (1)$$

with $\Sigma Q_i^2 = Q_0^2$ splits into three equivalent though interdependent eigenvalue equations along the vibrational mode coordinates $Q_i$:

$$[-(\eta^2/2M)\Sigma_i(\partial^2/\partial Q_i^2)+(d_b-d_c)(M\omega_n^2/b)Q_i^4]\Phi_i(Q_l)= E_i\Phi_i(Q_l) \qquad (2)$$

For any dipolar species to reorientate, each of its coupled mode coordinates $Q_i$ should undergo a hindered displacement in configurational space controlled by the respective potential energy profile. Given the formal populations $n_i$ along $Q_i$, the rate equations for relaxation of a reorientating system are:

$$dn_i/dt = -k_i n_i$$

where $k_i$ is the rate constant along $Q_i$, provided there is no population interchange between the configurational axes. We get

$$n_i(t) = n_i(0)\exp(-k_i t) \qquad (4)$$

For a system that has initially been distributed uniformly along all $Q_i$-axes, $n_i(0) = (1/3)n_0$, its axial time evolution will be determined by the individual potential energy profile along each $Q_i$. With that profile similar along each of the three axes we get

$$n(t) = \Sigma_i n_i(t) = (1/3)n_0 \Sigma_i \exp(-k_i t) = n_0 \exp(-k_{relax} t) \qquad (5)$$

i.e. $k_i = k_{relax}$ is independent of the axis, and the relaxation of that system may be described as if one-dimensional along an axis. For such systems, therefore, the relaxation coordinate may be assumed rectilinear which makes it unnecessary taking into account the centrifugal effects arising from the curvature during reorientation. In other cases rectilinearity may also be expected to provide a good approximation in a two-site approach to processes along a small-curvature relaxation coordinate.

### 2.2. Two-site rate

#### 2.2.1. Definition

While the reorientational motion of an off-centered impurity ion is actually controlled by a multiwell potential surface along the 'relaxation coordinate', a double-well (two-site) based analysis is usually found applicable, leading to $\tau_{rel} = (gk_{12})^{-1}$ where $k_{12}$ is the two-site relaxation rate and $g$ is an effective number of equivalent wells depending on the symmetry of dipole and lattice, as well as on the orientation of the external electric field.

The usual theoretical prediction for the relaxation rate $k_{12}$ is based on the multiphonon approach (MPA) [8]. Mathematically MPA rests on the time-dependent perturbation theory which defines the rate by means of Fermi's golden rule:

$$k_{12}(T) = (2\pi/\eta)\Sigma_{n1,n2} | F(n_1,T)|<j_2,n_2|H'|j_1,n_1>|^2 \delta(E_n - E_{n1}) \qquad (6)$$

where $\eta = h/2\pi$ while the sum is over the final states which conserve energy relative to the corresponding initial states, weighted by means of the thermal occupation probabilities $F(E_{n1},T)$. H' is the relaxation-driving part of the Hamiltonian composed of the nuclear kinetic-energy operator (the nonadiabaticity operator). A further step is Condon's approximation to factorize the matrix element into electronic and nuclear components

$$M_{12} = <j_2,n_2| H' |j_1,n_1> = <j_2,Q| H' |j_1,Q><n_2|n_1> \qquad (7)$$

Inasmuch as the transition probability is proportional to $|M_{12}|^2$, Condon's assumption effects factorization of that probability into electronic and nuclear parts. The electronic part

$$K_{12} = <j_2,Q| H' |j_1,Q> \qquad (8)$$

is assumed small to legitimize the use of the perturbation method. This confines the theory to transitions in which the initial and final electronic states $j_1$ and $j_2$ are only weakly coupled. Note that because of $Q_1 \neq Q_2$ (as a result of the electron-phonon interaction) the nuclear counterpart can be nonvanishing even though $n_1 \neq n_2$.

The reaction-rate approach (RRA) on the other hand is based on an occurrence-probability formula accounting for both classical and quantal effects [5]. Now the transition rate is given by:

$$k_{12}(T) = k(T)(k_BT/\eta)(Z_1^\#/Z_1)\exp(-E_B/k_BT) \qquad (9)$$

$$k(T) = \Sigma_n W_n(E_n)\exp(-\varepsilon_n\}/k_BT) \delta\varepsilon_n\}/k_BT) \qquad (10)$$

is a correction factor to the rate equation, presented otherwise in its conventional classical form. k(T), therefore, accounts for both nonadiabaticity and quantal effects. $Z_1$ is the complete partition function of the initial state, assumed to be in thermal equilibrium, $Z_1^\#$ is the initial-state partition function of the nonreactive accepting modes obtained by excluding the relaxation-promoting mode from the domain of all the modes entering into $Z_1$. $E_B$ is the barrier height (II.36) between the two neighboring reorientational sites at 1 and 2 along the relaxation coordinate. The motion along that coordinate is quantized with n, $E_n$ being the energy; $\varepsilon_n = E_n - E_B$ is the excess energy relative to the barrier peak, while $\delta\varepsilon_n = E_{n+1} - E_n$ is the separation between subsequent energy levels. $k_B$, h and T have their usual meanings.

Calculating the quantum correction k(T) often reduces to an one-dimensional problem which has been solved when the relaxation coordinate is separable dynamically from the domain of all mode coordinates. $W_n(E_n)$, the transition probability at energy $E_n$, is assumed to factorize in concert with Condon's approximation:

$$W_n(E_n) = W_{Ln}(E_n)W_{en}(E_n) \qquad (11)$$

into $W_L$, the probability for configurational rearrangement, and $W_e$, the probability for a change of the electronic state at the transition configuration $Q_C$ between the initial and final states (e.g.

$a_{1g}$ and $t_{1u}$) under the energy conservation condition. Transitions with $W_e = 1$ are called 'adiabatic', those with $W_e < 1$ are 'notadiabatic', while those with $W_e \ll 1$ are 'nonadiabatic'. Making next use of the assumed harmonicity of lattice vibrations, the promoting-mode contribution, factorized out of $Z_1$ under the separability condition, gives

$$Z_1^{\#}/Z_1 = 2\sinh(h\nu_r\}/2k_BT) \tag{12}$$

where $\nu_r$ ($\omega_r$) is the effective vibrational (angular) frequency along the relaxation coordinate. Inserting into (9) one obtains finally

$$k_{12}(T) = k(T)(2k_BT/h\nu)/\sinh(h\nu_r/2k_BT)\nu_r\exp(-E_b/k_BT) \tag{13}$$

in the harmonic mode approximation. It is noteworthy that eq.(13) predicts a non-vanishing zero-point rate of magnitude

$$k_{12}(0) = \nu_r W_{e0}(E_0) W_{L0}(E_0) \tag{14}$$

with $E_0 = (1/2) h \nu_r$.

In view of the strong anharmonicity of hindered rotation, however, equations (12) and (13) should be replaced by respective ones applicable to the particular case. For a hindered rotation, the eigenvalue spectrum being

$$E_{a/b,n} = (\eta^2/2I_A)a_n, b_n = (\eta^2/2I_A)[n^2+c_n(q)],$$

where $c_m(q)$ is the correction either to $a_m$ or to $b_m$ due to the hindering potential, we get

$$Z_1^{\#}/Z_1 = 1/\Sigma_{m=0}\exp(-\eta^2[m^2+c_m(q)]/2I_Ak_BT)$$

$$= \exp(\eta^2 c_0(q)/2I_Ak_BT) \times$$

$$\{1 + \Sigma_{m=1}^{\infty}\exp(-\eta^2[m^2+c_m(q)-c_0(q)]/2I_Ak_BT)\} \tag{15}$$

over the eigenspectrum of the particular rotating mode if separable from the domain of all rotational modes. For a formal comparison, the harmonic-oscillator term (12) obtains from (15) by substituting $\eta\omega_r$ for $\eta^2/2I_A$, n for $m^2$, and 1/2 for $c_m(q)$. The sum in the denominator may be computed numerically at not too large q, following the prescriptions of Part II.

The relaxation rate obtains by combining equations (9) & (15):

$$k_{12}(T) = k(T)(k_BT/\eta\nu_r)\exp(\eta^2 c_0(q)/2I_Ak_BT)\{1 + \Sigma_{m=1}^{\infty}\exp(-\eta^2[m^2+c_m(q)-c_0(q)]/2I_Ak_BT)\}\nu_r \times$$

$$\exp(-E_b/k_BT) \tag{16}$$

Insofar as $c_m(q)$ are but small corrections to the basic $m^2$ terms in $a_m$ and $b_m$, a zero-point rate of magnitude

$$k_{12}(0) = (\eta/4\pi I_A)[1+c_1(q)-c_0(q)]W_{e0}(E_0)W_{L0}(E_0) \tag{17}$$

is predicted at $E_0 = (\eta^2/2I_A)c_0(q)$, $\eta/2I_A$ playing the role of an effective rotational frequency.

### 2.2.2. Electron transfer probability

Available calculations of the electron-transfer term $W_e$ are largely based on Landau-Zener's quasiclassical approach though there also are quantum-mechanical justifications [5]. Introducing Landau's parameter

$$\gamma(E_n) = (E_{12}^2/4h\nu)E_R^{1/2} | E_n - E_C |^{-1/2} \qquad (18)$$

in which $E_C$ is the crossover energy, the electron-transfer probability for multiple overbarrier transitions at $E_n \gg E_B$ has been derived to be [12]

$$W_e(E_n) = 2[1 - \exp(-2\pi\gamma)]/[2 - \exp(-2\pi\gamma)] \qquad (19)$$

A nonadiabatic transfer $W_e = 4\pi\gamma \ll 1$ obtains at $\gamma \ll 1$. For underbarrier transitions $E_n \ll E_B$,

$$W_e(E_n) = 2\pi \gamma^{2\gamma-1} \exp(-2\gamma)/[\Gamma(\gamma)]^2 \qquad (20)$$

Again, $W_e = 2\pi\gamma \ll 1$ for a nonadiabatic transfer at $\gamma \ll 1$.

### 2.2.3. Configurational tunneling probability.

Following general arguments [5], the transition probability along the angular coordinate $\varphi$ reads

$$W_{if}(E_n) = 4\pi^2 |V_{fi}|^2 \sigma_i(E_n)\sigma_f(E_n) \qquad (21)$$

where the matrix element $V_{fi}$ is to be calculated using initial and final state wavefunctions $u_i$ and $u_f$, respectively, as

$$V_{fi} = (\eta^2/2I_A)[u_f^*(du_i/d\varphi) - u_i(du_f^*/d\varphi)]|_{\varphi=\varphi_c} \qquad (22)$$

Here $\sigma_i(E_n)$ and $\sigma_f(E_n)$ are the corresponding densities (DOS) of the initial and final states.

#### 2.2.3.1. General solution

The exact rotational eigenstates are constructed by Mathieu's functions $Y(\varphi)$, $u_i(\varphi) = Y_i(\varphi + \pi/4)$, $u_f(\varphi) = Y_f(\varphi - \pi/4)$, while the densities of states are both given by

$$\sigma(E_{a/b,n}) = dn/dE_{a/b,n} = (2I_A\eta^2)(dn/da_n, b_n) \qquad (23)$$

since the rotational eigenspectrum is $E_{a/b,n} = (\eta^2/2I_A)a_n, b_n$. Now using $Y_{i/f}(\varphi, q)$, equation (22) at the saddle point $\varphi = 0$ reads:

$$V_{if} = (-\eta^2/2I_A)\{Y_f(-\pi/4, q)[dY_i(\varphi, q)/d\varphi]|_{\varphi=\pi/4} - Y_i(\pi/4, q)[dY_f(\varphi, q)/d\varphi]|_{\varphi=-\pi/4}\} \qquad (24)$$

inasmuch as for $\varphi$ and $q$ real $Y_{i/f}(\varphi, q)$ is a real function of $\varphi$. Series expansions of Mathieu's functions $ce_m(z,q)$ and $se_m(z,q)$ are available (see Part II) and will be applied to constructing rotational states and deriving transition probabilities $W_{fi}$.

Using the expansions in q, we calculate finite-valued saddle-point functions at $z = \pm \pi/2$, $\varphi = \pm \pi/4$), as in Part II. We see that each of the periodic functions $ce_m(z,q)$ and $se_m(z,q)$ is either itself vanishing or has a vanishing derivative at $z = \pi/2$. For this reason, $ce_m(z,q)$ and $se_m(z,q)$ are not by themselves the appropriate rotational eigenstates but such may be constructed as linear combinations of the basic functions.

For reasons transparent from the discussion in Part II, we consider the following linear combinations of Mathieu's functions $lc_m^+(z,q) = ce_m(z,q) + ce_{m+1}(z,q)$ and $ls_m^+(z,q) = se_m(z,q) + se_{m+1}(z,q)$, Equation (24) is redefined in terms of $z = 2\varphi$ to give

$$V_{if,cm} = 2(-\eta^2/2I_A)\, 2lc_m^+(\pi/2,q)[(d/dz)lc_m^+(z,q)]|_{z=\pi/2} =$$

$$2(-\eta^2/2I_A)(d/dz)[lc_m^+(z,q)]^2|_{z=\pi/2}$$

or, alternatively,

$$V_{if,sm} = 2(-\eta^2/2I_A)\, 2ls_m^+(\pi/2,q)[(d/dz)ls_m^+(z,q)]|_{z=\pi/2} = 2(-\eta^2/2I_A)(d/dz)[ls_m^+(z,q)]^2|_{z=\pi/2}$$

The transition matrix element $V_{if}$ is now finite since for any two consecutive quantum numbers $n = m, m+1$ either a component function ($ce_n(z,q)$ or $se_n(z,q)$) or its derivative is finite at $z = \pi/2$. As examples, we get by even linear combination

$$V_{fi,c0} = 2(-\eta^2/2I_A)\, ce_0(\pi/2,q)[dce_1(\pi/2,q)/dz]$$

$$V_{fi,c1} = 2(-\eta^2/2I_A)\, ce_2(\pi/2,q)[dce_1(\pi/2,q)/dz]$$

$$V_{fi,c2} = 2(-\eta^2/2I_A)\, ce_2(\pi/2,q)[dce_3(\pi/2,q)/dz]$$

$$V_{fi,c3} = 2(-\eta^2/2I_A)\, ce_4(\pi/2,q)[dce_3(\pi/2,q)/dz]$$

$$V_{fi,c4} = 2(-\eta^2/2I_A)\, ce_4(\pi/2,q)[dce_5(\pi/2,q)/dz],$$

etc. and also by odd linear combination

$$V_{fi,s1} = 2(-\eta^2/2I_A)\, se_1(\pi/2,q)[dse_2(\pi/2,q)/dz]$$

$$V_{fi,s2} = 2(-\eta^2/2I_A)\, se_3(\pi/2,q)[dse_2(\pi/2,q)/dz]$$

$$V_{fi,s3} = 2(-\eta^2/2I_A)\, se_3(\pi/2,q)[dse_4(\pi/2,q)/dz]$$

$$V_{fi,s4} = 2(-\eta^2/2I_A)\, se_5(\pi/2,q)[dse_4(\pi/2,q)/dz]$$

Referring to the discussion in Part II again, we see that type (22) intraband transitions are only nonvanishing if these occur in bands along the upper-sign branch of (39), such as ones described by the linear combinations

$$u(z,q) = (1/2)[ce_{m-1}(z,q) + ce_m(z,q)] = (1/2)lc_{m-1}^+(z,q) \text{ for } m \text{ odd}$$

$$u(z,q) = (1/2)[se_{m-1}(z,q) + se_m(z,q)] = (1/2)ls_{m-1}^+(z,q) \text{ for } m \text{ even}$$

whereas lower-sign branch bands, such as ones described by the linear combinations

$u(z,q) = \frac{1}{2}[ce_{m-1}(z,q) + se_m(z,q)]$ for all m,

do not promote any such transitions.

We first illustrate this statement for the lowest energy bands at "negative q" and "positive q", respectively:

$u_f^*(du_i/dz) - u_i^*(du_f/dz)|_{z=\pi/2} = \frac{1}{2}[ce_0(\pi/2,q) + ce_1(\pi/2,q)]\{[dce_0(\pi/2,q)/dz] + [dce_1(\pi/2,q)/dz]\} =$

$\frac{1}{2} ce_0(\pi/2,q)[dce_1(\pi/2,q)/dz]$, q<0

$u_f^*(du_i/dz) - u_i^*(du_f/dz)|_{z=\pi/2} = \frac{1}{2} ce_0(\pi/2,q)[dce_1(\pi/2,q)/dz] - \frac{1}{2} se_1(\pi/2,q)[dse_1(\pi/2,q)/dz] = 0$, q>0

etc. (cf. Fig.II.2). The former generalizes straightforwardly to

$u_f^*(du_i/dz) - u_i^*(du_f/dz)|_{z=\pi/2} =$

$\frac{1}{2}[ce_{m-1}(\pi/2,q) + ce_m(\pi/2,q)]\{[dce_{m-1}(\pi/2,q)/dz] + [dce_m(\pi/2,q)/dz]\} =$

$\frac{1}{4} d[ce_{m-1}(z,q) + ce_m(z,q)]^2/dz|_{z=\pi/2}$

for odd m = 1,3,5,... and to

$u_f^*(du_i/dz) - u_i^*(du_f/dz)|_{z=\pi/2} =$

$-\frac{1}{2}[se_{m-1}(\pi/2,q) + se_m(\pi/2,q)]\{[dse_{m-1}(\pi/2,q)/dz] + [dse_m(\pi/2,q)/dz]\} =$

$-\frac{1}{4} d[se_{m-1}(z,q) + se_m(z,q)]^2/dz|_{z=\pi/2}$

for even m = 2,4,6,... The relevant transition probabilities are

$W_{if}(E_m) = 4\pi^2 N_m^{-4} [2dm/d(a_{m-1}+a_m)]^2 [ce_{m-1}(\pi/2,q) + ce_m(\pi/2,q)]^2 \times$

$\{[dce_{m-1}(z,q)/dz]|_{z=\pi/2} + [dce_m(z,q)/dz]|_{z=\pi/2}\}^2 =$

$[2dm/d(a_{m-1}+a_m)]^2 \{d[ce_{m-1}(z,q) + ce_m(z,q)]^2/dz|_{z=\pi/2}\}^2$ (m = 1,3,5,..)

$W_{if}(E_m) = 4\pi^2 N_m^{-4} [2dm/d(b_{m-1}+b_m)]^2 [se_{m-1}(\pi/2,q) + se_m(\pi/2,q)]^2 \times$

$\{[dse_{m-1}(z,q)/dz]|_{z=\pi/2} + [dse_m(z,q)/dz]|_{z=\pi/2}\}^2 =$

$[2dm/d(b_{m-1}+b_m)]^2 \{d[se_{m-1}(z,q) + se_m(z,q)]^2/dz|_{z=\pi/2}\}^2$ (m = 2,4,6,..)     (25)

$W_{if}$ are constructed by linear combinations $\frac{1}{2} lc_{m-1}^+(z,q)$ and $\frac{1}{2} ls_{m-1}^+(z,q)$ of normalized eigenfunctions $N_m ce_m(z,q)$ and $N_m se_m(z,q)$ with $N_m = \pi^{-1/2}$, the linear combinations corresponding to the energy eigenvalues $E_m = (\eta^2/4I_A)(a_{m-1}+a_m)$ for m = 1,3,5,... odd and $E_m = (\eta^2/4I_A)(b_{m-1}+b_m)$ for m = 2,4,6,... even, respectively.

Strictly speaking, the above linear combination eigenstates corresponding to energy eigenvalues in the middle of the allowed bands, they do not adequately account for the interior of these bands. To improve the description, we make the following proposition: We attach an integer n to number a band where n is odd for $(a_{n-1}, a_n)$ and even for $(b_{n-1}, b_n)$, and let m be a running number $0 \leq m \leq 1$. In so far as the eigenfunctions $ce_m(z,q)$ and $se_m(z,q)$ describing intraband states at noninteger m are not available, we form intraband states by way of linear combinations of band-edge states:

$$ce^n_m(z,q) = (1-m)ce_{n-1}(z,q) + m\, ce_n(z,q) \quad (n\ \text{odd})$$

$$se^n_m(z,q) = (1-m)se_{n-1}(z,q) + m\, se_n(z,q) \quad (n\ \text{even}) \tag{26}$$

with intraband eigenvalues

$$E^n_m(q) = (1-m)E_{n-1} + mE_n = (\eta^2/2I)a^n_m(q), \quad a^n_m(q) = (1-m)a_{n-1}(q) + ma_n(q)$$

(n=1,3,5,...odd)

$$E^n_m(q) = (1-m)E_{n-1} + mE_n = (\eta^2/2I)b^n_m(q), \quad b^n_m(q) = (1-m)b_{n-1}(q) + mb_n(q)$$

(n=2,4,6,...even) \qquad (27)

respectively. Using the so-constructed intraband states, we redefine the transition probabilities

$$W_{Ln}(E^n_m) = (2\pi)^2 |V^n_m(q)|^2 (dm/dE^n_m)^2$$

so as to incorporate

$$V^n_m(q) = -(2\eta^2/I\pi)ce^n_m(z,q)[dce^n_m(z,q)/dz]|_{z=\pi/2} = -(2\eta^2/I\pi)(1-m)ce_{n-1}(z,q)[mdce_n(z,q)/dz]|_{z=\pi/2}$$

(n odd)

$$V^n_m(q) = -(2\eta^2/I\pi)se^n_m(z,q)[dse^n_m(z,q)/dz]|_{z=\pi/2} = -(2\eta^2/I\pi)(1-m)se_{n-1}(z,q)[mdse_n(z,q)/dz]|_{z=\pi/2}$$

(n even) \qquad (28)

We get accordingly

$$W_{Ln}(E^n_m) = 64 \left| (1-m)ce_{n-1}(z,q)[mdce_n(z,q)/dz] \right|^2_{z=\pi/2} (dm/da^n_m)^2$$

$$= 64 \left| (1-m)se_{n-1}(z,q)[mdse_n(z,q)/dz] \right|^2_{z=\pi/2} (dm/db^n_m)^2 \tag{29}$$

The above probabilities (29) are maximum in the middle of a band at m = 1/2 ($W_{Ln}^{max}$) and vanish at the band edges at m=0 and m=1. To work out an expression feasible for practical calculations eqn.(29) should be normalized to unity.

In cases where Mathieu's functions can be approximated for by free-rotor eigenstates e.g. $Y_m(\varphi,0) = \pi^{-1/2}\cos(m\varphi)$ we get $V_{fi}(E_n) \sim (\eta^2/2I\pi)m[-\cos[m(\varphi-\pi/4)]\sin[m(\varphi+\pi/4)] + \cos[m(\varphi+\pi/4)]\sin$

$[m(\varphi+\pi/4)]|_{\varphi=0}$ which is equal to $\pm(\eta^2/2I\pi)m$ for m odd and to 0 for m even. If we set $a_m = m^2$ leading to $\sigma(E_m) = (I/\eta^2)(1/m)$ we obtain $W_{if}(E_m) = 4\pi^2(\eta^2/2I\pi)^2m^2 (I/\eta^2)^2(1/m)^2 = 1$ for m odd and $W_{if}(E_m) = 0$ for m even. It implies that the configurational probability of a free rotor is energy-independent, as it should. However if we set $a_m = a_m(q)$ leading to $\sigma(E_n) = (2I/\eta^2)[dn/da_n(q)]$ we obtain $W_{if}(E_m) = 4\pi^2(\eta^2 n/2I\pi)^2 (2I/\eta^2)^2 (dn/da_n)^2 = 4n^2 (dn/da_n)^2$ which is attributed to quasi-free rotations well above the barrier top.

### 2.2.3.2. Well-bottom solution

A configurational-tunneling probability has otherwise been calculated quantum-mechanically for parabolic wells with $E_{12} \ll E_B$ using harmonic-oscillator wavefunctions to derive $V_{fi}$ by means of equation (22), though in other cases quasiclassical techniques have been applied to get a result. For underbarrier transitions at $E_n \ll E_B$ and isoenergetic parabolic wells whose bottoms lie at the same energy as in the reorientation under consideration, Christov's quantum-mechanical expressions have been derived [5]:

$$W_L(E_n) = \pi[F_{nn}(\xi_0,\xi_C)/2^n n!]^2 \exp(-E_R/h\nu) \qquad (30)$$

where $n = n_1 = n_2$ is the vibronic quantum number in the initial (final) electronic state, $\xi_0 = \xi_2 - \xi_1$ is the interwell separation along the relaxation coordinate, and

$$F_{nn}(\xi_0,\xi_C) = \xi_0 H_n(\xi_C) H_n(\xi_C-\xi_0) - 2nH_{n-1}(\xi_C)H_{n-1}(\xi_C-\xi_0) + 2nH_n(\xi_C)H_{n-1}(\xi_C-\xi_0) \qquad (31)$$

with $H_n(q)$ standing for Hermite's polynomials. For overbarrier transitions at $E_n \gg E_B$ one sets $W_L = 1$. One is to apply specific techniques to deal with transitions near the barrier top $E_n \sim E_B$. We remind that $\xi = (M\omega_{ren}/\eta)^{1/2}Q$ or $\xi = (I\omega_{ren}/\eta)^{1/2}\varphi$ stand for the scaled vibrational or rotational coordinates. The well-bottom relaxation rate obtains through insertion in equations (10)&(13).

### 2.2.4. Reorientation rates

### 2.2.4.1. Mathieu rates

We shall next use the intraband transition probabilities to derive a practical expression for a relaxation rate pertinent to a system of allowed bands. Integrating over a rotational energy band and summing up for all the bands, we get a transition rate:

$$k_{12}(T) = (Z^\#_1/Z_1) \Sigma_{n=1}^{\infty} \int_{E_{n-1}}^{E_n} W_{en}(E^n_m) W_{LNn}(E^n_m)\exp(-E^n_m/k_BT)dE^n_m/\eta \qquad (32)$$

where $W_{LNn}$ are the normalized configurational probabilities

$$W_{LNn}(E^n_m) = 64 \times |(1-m)ce_{n-1}(z,q)[mdce_n(z,q)/dz]|^2_{z=\pi/2} (dm/da^n_m)^2$$

$$= 64 \times |(1-m)se_{n-1}(z,q)[mdse_n(z,q)/dz]|^2_{z=\pi/2} (dm/db^n_m)^2$$

where the normalization factor is defined by

$$N^{-1} = 2\Sigma_{n=1}^{\infty} \int_0^1 W_{Ln}(E^n_m)dm = 128 \Sigma_{n=1}^{\infty} \int_0^1 dm[m(1-m)]^2 \times$$

$$\begin{Bmatrix} |ce_{n-1}(z,q)[dce_n(z,q)/dz]|^2_{z=\pi/2}(dm/da^n_m)^2 \\ |se_{n-1}(z,q)[dse_n(z,q)/dz]|^2_{z=\pi/2}(dm/db^n_m)^2 \end{Bmatrix}$$

The DOS of a hindered rotator will be derived from equation (27):

$dm/da_m(q) = 1/[a_n(q)-a_n(q)]$, $dm/b_m(q) = 1/[b_n(q)-b_{n-1}(q)]$.

(For comparison, the DOS of a free rotator is $dm/da_m(0) = 1/2m$.) We extend the definition of $E^n_m$ to negative m so as

$E^n_m = E_{n-1} + m(E_n-E_{n-1})$ $(0 < m < 1)$

$E^n_m = E_n + m(E_n-E_{n-1})$ $(-1 < m < 0)$

and formulate the rate accordingly

$k_{12}(T) = (16\eta/\pi I)N(Z_1^{\#}/Z_1) \times$

$\Sigma_{n=1}^{\infty} \{ \exp(-E_{n-1}/k_BT) \int_0^1 dm\, W_{en}(E^n_m)\exp(-m[E_n-E_{n-1}]/k_BT)[(1-m)m]^2$

$+ \exp(-E_n/k_BT) \int_{-1}^0 dm\, W_{en}(E^n_m)\exp(-m[E_n-E_{n-1}]/k_BT)[(1-m)m]^2 \} \times$

$$\begin{Bmatrix} [a_n(q)-a_{n-1}(q)]^{-1}|ce_{n-1}(z,q)[dce_n(z,q)/dz]|^2_{z=\pi/2} \\ [b_n(q)-b_{n-1}(q)]^{-1}|se_{n-1}(z,q)[dse_n(z,q)/dz]|^2_{z=\pi/2} \end{Bmatrix} \quad (33)$$

For $W_{en}(E^n_m)$ constant as in an adiabatic process, the integration over m can be carried out in (33). The result splits the rate into two parts, as follows:

$k_{12}(T)=(16\eta/\pi I)N(Z_1^{\#}/Z_1)\Sigma_{n=1}^{\infty} W_{en}(E_m)(2/g_n)\{-(1/g_n^2)[1+(6/g_n)+(12/g_n^2)]\exp(-E_n/kBT) +$

$[2-(6/g_n)+(13/g_n^2)-(18/g_n^3)+(12/g_n^4)]\exp(-E_{n-1}/k_BT)\} \times$

$$\begin{Bmatrix} [a_n(q)-a_{n-1}(q)]^{-1}|ce_{n-1}(z,q)[dce_n(z,q)/dz]|^2_{z=\pi/2} \\ [b_n(q)-b_{n-1}(q)]^{-1}|se_{n-1}(z,q)[dse_n(z,q)/dz]|^2_{z=\pi/2} \end{Bmatrix} \quad (34)$$

where $g_n =(E_n-E_{n-1})/k_BT$. The partition function derives from

$(Z^{\#}_1/Z_1)^{-1} = \Sigma_{n=1}\infty \int_{E_{n-1}}^{E_n} \exp(-E^n_m(q)/k_BT)(dm/dE^n_m)dE^n_m$

$= \Sigma_{n=1}^{\infty} \{\exp(-E_{n-1}/k_BT) \int_0^1 \exp(-m[E_n-E_{n-1}]/k_BT)dm +$

$\exp(-E_n/k_BT)\int_{-1}^0 \exp(-m[E_n-E_{n-1}]/k_BT)dm\}$

$= \Sigma_{n=1}^{\infty}(1/g_n)\{[1-\exp(-g_n)]\exp(-E_{n-1}/k_BT) + [\exp(g_n)-1]\exp(-E_n/k_BT)\}$

$= \Sigma_{n=1}^{\infty} (2/g_n)[1-\exp(-g_n)]\exp(-E_{n-1}/k_BT)$  (35)

At low temperature, $k_{12}(T)$ varies mainly as $(4/g_1)\exp(-E_0/k_BT)$, while $(Z^{\#}_1/Z_1)^{-1}$ does so as $(2/g_1)\exp(-E_0/k_BT)$. Combining we get a zero-point rate

$k_{12}(0) = W_e (32\eta/\pi I) N[a_1(q)-a_0(q)]^{-1} |ce_0(z,q)[dce_1(z,q)/dz]|^2_{z=\pi/2}$  (36)

The zero-point rate accounts for the contribution of the lowest allowed band only. We remind that the particular form of equation (39) was obtained assuming a constant DOS over each rotational band. Here and above the normalization constant N given by

$N^{-1} = (128/30 \Sigma_{n=1}^{\infty} \times$

$\{ |ce_{n-1}(z,q)[dce_n(z,q)/dz]|^2_{z=\pi/2} (a_n-a_{n-1})^{-2}$

$|ce_{n-1}(z,q)[dce_n(z,q)/dz]|^2_{z=\pi/2} (a_n-a_{n-1})^{-2}$
.

is derived through the doubled integration over the m > 0 states.

### 2.2.4.2. Harmonic rates

It should be stressed that the harmonic-oscillator based rate formulae apply to a system of discrete or nearly discrete energy levels, such as the rotational bands at large q. At small $V_{if}$, the harmonic rate obtains by combining equations (10), (11), (13), (30), and (31). At large $V_{if}$, quasiclassical techniques may be advised if no quantum-mechanical formulae are available for the configurational terms. However, a quantum-mechanical formula should always be applied to the relaxation from the ground-state vibronic level. Accordingly, RRA predicts a constant zero-point harmonic-oscillator rate at low temperature given by

$k_{12}(0) = W_{e0}(E_0)(E_{RII}/\eta)\exp(-E_{RII}/\eta\omega_{renII})$  (37)

at $E_0 = (1/2)\eta\omega_{renII}$ to be compared with the Mathieu rate (36).

### 2.2.4.3. Experimental checkup

The temperature dependence of a two-site linear rate pertinent to $Li^+$ in KCl has been discussed earlier [10] compared with the experimental paraelectric resonance data [7]. These data are interesting in that they suggest three different though parallel relaxation channels below 5K, along <100>, <110> and <111>, respectively, which all converge to another single process above 12K. The three channels below 5K are apparently different projections of the basic <111> rate along the corresponding electric field orientations which projections decrease in magnitude from <111> to <110> to <100>, as they should. Their convergence above 12K may result from the increased smearing of the $Li^+$ ion within the off-centered ellipsoid during the response time of experiment. Smearing may also affect the apparent barrier height proportional to the observed temperature slope. The slopes are roughly 7 meV below 5K and 70 meV above 12K.

It is essential that RRA predicts constant zero-point rates at the lowest temperatures, namely, renormalized-linear given by equation (37) for the ground-state harmonic level and nonlinear by equation (36) for the lowest rotational band, respectively. Nevertheless, no constant zero-point rates are exhibited in the experimental rate versus temperature plots.

The zero-point rate (the order of $10^9$ s$^{-1}$, if at all) may have remained unseen experimentally at temperatures below 2K. Alternatively, the involvement of strong vertical tunneling one-phonon processes below 5K has been suggested and indeed these are not covered by RRA which accounts for horizontal tunneling energy conserving transitions only which conserve the phonon occupation number. Indeed, it seems likely that one-phonon processes may be decisive at low temperatures. Nevertheless, zero-point rates are revealed in experiments with off-center ions such as Ag$^+$ in RbBr [11] and F$^-$ in NaBr, KI, RbI [10].

Tables I and II compare relaxation rate data, viz. zero-point rates as derived in the harmonic bottom-well approximation and by Mathieu's mathematics, respectively. While the harmonic solution adapts to the well bottom, it may not do so to the barrier top [6]. This has also been shown elsewhere using KCl data where the parabolic bottom-fitting potential has been compared with the actual cosine-type. Consequently, the Mathieu solutions may be expected to do better especially in cases where the harmonic approximation fails. The Tables display Mathieu rates generally superior to the harmonic rates.

Tables III and IV present band-edge Mathieu eigenvalues and eigenstates of an impurity in KCl, isolated and near an F center, respectively. The maximal configuration probabilities at mid-band energies are also computed according to

$W_{LN}^{nmax}(E^n_{1/2}) = N \times$

$\begin{cases} |2ce_{n-1}(z,|q|)[dse_n(z,|q|)/dz]|^2 _{z=0} (a_n-a_{n-1})^{-2} \\ |2ce_{n-1}(z,|q|)[dse_n(z,|q|)/dz]|^2 _{z=0} (b_n-b_{n-1})^{-2} \end{cases}$

$1/N = (128/30) \Sigma_{n=1}^\infty \times$

$\begin{cases} |ce_{n-1}(z,|q|)[dse_n(z,|q|)/dz]|^2 _{z=0}(a_n-a_{n-1})^{-2} \\ |ce_{n-1}(z,|q|)[dse_n(z,|q|)/dz]|^2 _{z=0}(a_n-a_{n-1})^{-2} \end{cases}$

for n odd and even, respectively. These probabilities are seen to decrease slightly as the band number n is increased which is not surprising, since the corresponding bands grow wider while DOS decreases in the inverse proportion.

Finally Figure 2 shows a comparison of Christov- and Mathieu-based 2D relaxation times, as computed using equation (33) with integrals over m calculated numerically, with the experimental paraelectric data on Li$^+$ in KCl [7]. The resulting temperature dependences of the relaxation times exhibit zero-point rates followed smoothly by Arrhenius branches at higher temperatures. Clearly the 2D time goes somewhat closer to the experimental points, though this does not solve the controversy over the lower temperature behavior. A harmonic rate fit to the same data has been reported earlier [10].

3. Conclusion

We have presented a theory for the reorientation of off-center impurity ions in crystals which combines the reaction-rate approach to the relaxation times with the periodic solutions and allowed rotational bands of Mathieu's eigenvalue equation. Both configurational transition probabilities and thereby relaxation rates are derived and compared with ones based on the traditional well-bottom adapted harmonic approximation. An obvious prediction being that reorientation is only possible in rotational energy bands at the lower adiabatic branch, the impurity will remain frozen in the initial reorientational site should its energy be lifted to an excited electronic state by optical absorption.

Acknowledgements. Thanks are due to Drs. Vesselin Krastev and Peter Sharlandjiev for help in providing a great deal of bibliographic material for this study.

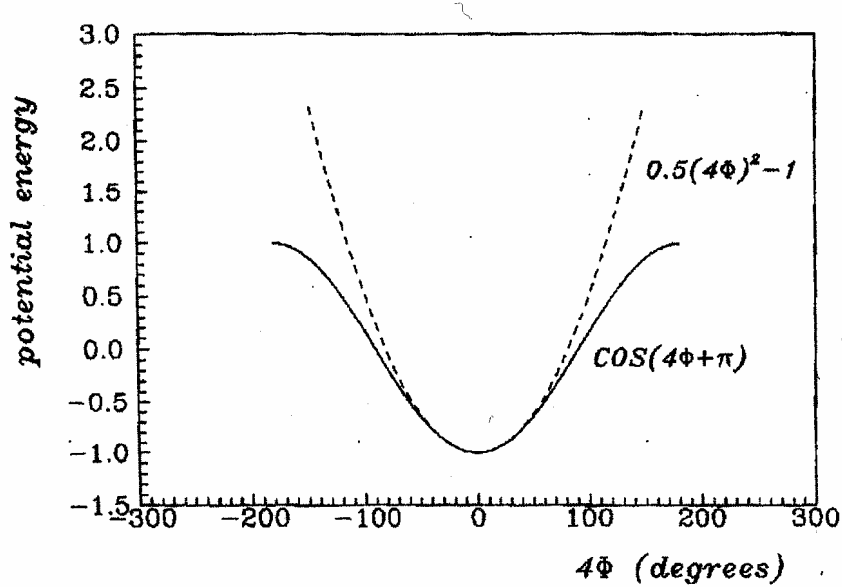

Figure 1: Comparing the well-bottom fitting parabolic potential with the actual cosine-type potential using Table I KCl data. It shows the applicability limits of the harmonic-oscillator model.

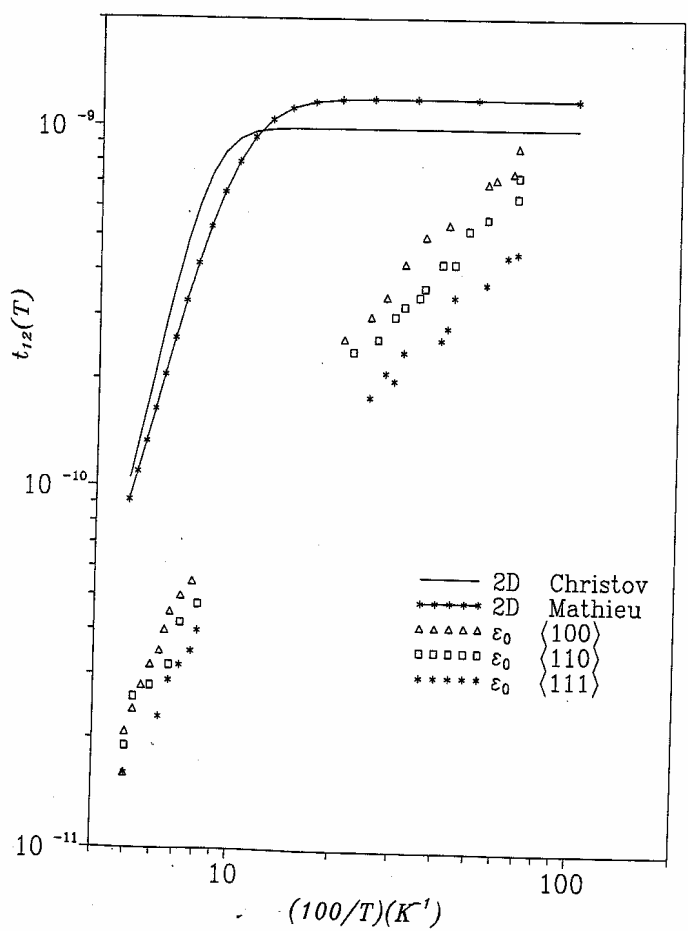

Figure 2: Comparing the temperature dependences of Mathieu and well-bottom (WB) based 2D relaxation times $\tau_{12}(T) = k_{12}(T)^{-1}$ of $Li^+$ in KCl, as calculated using the tabulated data, with experimental paraelectric data from Ref. [7]. The theoretical points are connected by a solid lines, and the experimental points taken at three different electric field polarizations are marked by symbols as follows: [100]-triangles, [110]-squares, [111]-stars. The large deviation of the experimental points from the theoretical lines between 1 to 10 K is due to one-phonon processes not accounted for presently. See Ref. [14] for greater details on the one-phonon rates.

Table I

Relaxation parameters for off-center $Li^+$

(Isolated impurity)[a]

|  | Christov's | linear | rates | Mathieu's | nonlinear | rates |
|---|---|---|---|---|---|---|
| Host | Relaxation Energy $E_{RII}$ (eV) | Phonon Energy $h\nu_{renII}$ (meV) | Adiabatic Rate at 0K $k_{12}(0)/W_e$ (s$^{-1}$) | Mathieu's Parameter -q | Mathieu's Lowest Energy (meV) | Mathieu's Rate at 0K $k_{12}(0)/W_e$ (s$^{-1}$) |
| LiF | 9.872 | 44.6 | « 1 | 2012 |  |  |
| NaF | 2.939 | 27.5 | « 1 | 468 |  |  |
| KF | 2.213 | 20.9 | « 1 | 458 |  |  |
| RbF | 4.633 | 22.3 | « 1 | 1777 |  |  |
| LiCl | 0.138 | 9.2 | 6.1×10$^7$ | 9 | 2.2 | 8.8×10$^8$ |
| " | 3.281 | 20.2 | « 1 | 1076 |  |  |
| NaCl | 0.585 | 12.7 | « 1 | 87 | 3.2 |  |
| KCl | 0.146 | 7.4 | 6.2×10$^5$ | 16 | 1.8 | 6.9×10$^6$ |
| RbCl | 0.243 | 7.6 | 5.0 | 41 | 1.9 |  |
| NaBr | 0.123 | 5.7 | 6.8×10$^4$ | 19 | 0.3 |  |
| KBr | 0.010 | 2.6 | 3.5×10$^{11}$ | 0.6 | 0.4 | 3.2×10$^{11}$ |
| RbBr | 0.029 | 3.1 | 4.3×10$^9$ | 3.5 | 0.8 | 1.6×10$^{10}$ |
| NaI | 0.005 | 2.0 | 6.7×10$^{11}$ | 0.2 | 0.2 | 1.1×10$^{12}$ |

[a] Based on $E_{RII} = \pi^2 E_{BII}$ with barrier data from Part II. Zero-point rate based on eq. (37): $k_{12} = W_{e0}(E_0)\nu_{renII}(E_{RII}/h\nu_{renII})\exp(-E_{RII}/h\nu_{renII})$. Also $C_- - E_{Lmin} = \frac{1}{2} E_{BII}$. Mathieu's zero-point rate is from eq.(36): $k_{12}(0) = (64\eta/\pi I)N[a_1-a_0]^{-1}|ce_0(z,q)ce_1(z,q)'|W_{em}(E_m)$ assuming $W_{em}(E_m) = 1$. The $a_n$ data are from Ref. [12] Tables. The Mathieu functions are from II and Ref. [13].

Table II

Relaxation parameters for off-center Li$^+$

(Impurity at F center)[b]

|  | Christov's | linear | rates | Mathieu's | nonlinear | rates |
|---|---|---|---|---|---|---|
| Host | Relaxation Energy $E_{RII}$ (eV) | Phonon Energy $h\nu_{renII}$ (meV) | Adiabatic Rate at 0K $k_{12}(0)/W_e$ (s$^{-1}$) | Mathieu's Parameter -q | Mathieu's Lowest Energy (meV) | Mathieu's Rate at 0K $k_{12}(0)/W_e$ (s$^{-1}$) |
| LiF | 9.305 | 53.8 | « 1 | 1227 |  |  |
| NaF | 2.702 | 33.0 | « 1 | 275 |  |  |
| KF | 2.042 | 25.2 | « 1 | 270 |  |  |
| RbF | 4.448 | 27.0 | « 1 | 1115 |  |  |
| LiCl | 2.934 | 24.1 | « 1 | 606 |  |  |
| NaCl | 0.496 | 14.9 | 2.6 | 46 | 3.7 |  |
| KCl | 0.098 | 8.2 | 1.0×10$^9$ | 5.8 | 1.9 | 3.1×10$^9$ |
| RbCl | 0.192 | 8.8 | 9.6×10$^4$ | 19.6 | 2.1 |  |
| NaBr | 0.063 | 5.9 | 2.1×10$^9$ | 4.7 | 1.4 | 5.4×10$^9$ |
| KBr | 0.001 | 1.7 | 3.8×10$^{11}$ | 0.0 | 0.0 | 1.7×10$^{12}$ |
| RbBr | 0.012 | 3.0 | 3.6×10$^{11}$ | 0.6 | 0.5 | 3.6×10$^{11}$ |

[b] Based on $E_{RIIA} = \pi^2 E_{BIIA}$ with barrier data from Part II. Zero-point rate based on eq. (37): $k_{12A} = W_{e0}(E_0)\nu_{renIIA}(E_{RIIA}/h\nu_{renIIA})\exp(-E_{RIIA}/h\nu_{renIIA})$. Also $C_- - E_{Lmin} = \frac{1}{2} E_{BIIA}$. Mathieu's zero-point rate is from eq.(36): $k_{12}(0) = (64\eta/\pi I)N[a_1-a_0]^{-1}|ce_0(z,q)ce_1(z,q)'|W_{em}(E_m)$ assuming $W_{em}(E_m) = 1$. The $a_n$ data are from Ref. [12] Tables. The Mathieu functions are from II and Ref. [13].

Table III

Mathieu's calculations for KCl[c]

(Isolated Impurity)[a]

| n | 0 | 1 | 2 | 3 | 4 | 5 |
|---|---|---|---|---|---|---|
| $a_n$ | -24.10 | -24.10 | 4.44 | 4.46 | 25.96 | 28.12 |
| $(\eta^2/2I)a_n$ | -5.61 | -5.61 | 1.03 | 1.04 | 6.05 | 6.6 |
| $b_{n+1}$ | -9.22 | -9.22 | 16.50 | 16.84 | 32.84 | 39.28 |
| $(\eta^2/2I)b_{n+1}$ | -2.15 | -2.15 | 3.84 | 3.92 | 7.65 | 9.15 |
| $E_{an}-E_{Lmin}$ | 1.79 | 1.79 | 8.43 | 8.44 | 13.45 | 13.95 |
| $E_{bn+1}-E_{Lmin}$ | 5.25 | 5.25 | 11.24 | 11.32 | 15.05 | 16.55 |
| $ce_n(z,q)$ | 0.00 | 0.01 | 0.06 | 0.26 | 0.71 | 1.15 |
| $se_{n+1}(z,q)$ | 0.01 | 0.08 | 0.33 | 1.00 | 2.22 | 3.77 |
| $W_{LNan-1nmax}$ | 0.20 | | 0.16 | | 0.15 | |
| $W_{LNbnn+1max}$ | 0.16 | | 0.16 | | 0.13 | |

[c] All energy units are in meV. Mathieu parameter $q = -15.91$, rotational energy quantum $\eta^2/2I = 0.23$, well bottom energy shift $C_- - E_{Lmin} = 7.4$, rotational energy $E_{an} = (\eta^2/2I)a_n + C_-$, $E_{bn} = (\eta^2/2I)b_n + C_-$, renormalized phonon quantum $h\nu_{renII} = 7.42$. $W_{LNan-1nmax} = N|2ce_{n-1}(z,q) \times ce_n'(z,q)|^2(a_n-a_{n-1})^{-2}$, $W_{LNbnn+1max} = N|2se_{n-1}(z,q)se_n'(z,q)|^2(b_n-b_{n-1})^{-2}$. The electron transfer $W_e(E_m)$ was set to be 1. The Mathieu functions are from Ref. [13] at $q = 16$. The normalization factor is $N = 6.97\times10^{-2}$.

Table IV

Mathieu's calculations for KCl

(Impurity at F Center)[d]

| n | 0 | 1 | 2 | 3 | 4 | 5 |
|---|---|---|---|---|---|---|
| $a_n$ | -7.07 | -7.06 | 7.81 | 9.16 | 17.57 | 25.67 |
| $(\eta^2/2I)a_n$ | -3.02 | -3.02 | 3.34 | 3.91 | 7.51 | 10.97 |
| $b_{n+1}$ | 1.35 | 1.50 | 12.29 | 16.81 | 25.76 | 36.48 |
| $(\eta^2/2I)b_{n+1}$ | 0.58 | 0.64 | 5.25 | 7.18 | 11.01 | 15.59 |
| $E_{an}-E_{Lmin}$ | 1.93 | 1.93 | 8.29 | 8.86 | 12.45 | 15.92 |
| $E_{bn+1}-E_{Lmin}$ | 5.53 | 5.59 | 10.20 | 12.13 | 15.96 | 20.54 |
| $ce_n(z,q)$ | 0.03 | 0.18 | 0.61 | 1.05 | 1.21 | 1.16 |
| $se_{n+1}(z,q)$ | 0.13 | 0.59 | 1.55 | 2.85 | 4.17 | 5.37 |
| $W_{LNan-1nmax}$ | 0.17 | | 0.18 | | 0.14 | |
| $W_{LNbnn+1max}$ | 0.18 | | 0.16 | | 0.12 | |

[d] All energy units are in meV. Mathieu parameter $q = -5.81$, rotational energy quantum $\eta^2/2I = 0.43$, well bottom energy shift $C_- - E_{Lmin} = 4.95$, rotational energy $E_{an} = (\eta^2/2I)a_n + C_-$, $E_{bn} = (\eta^2/2I)b_n + C_-$, renormalized phonon quantum $h\nu_{renII} = 8.22$. $W_{LNan-1nmax} = N|2ce_{n-1}(z,q)ce_n'(z,q)|^2 \times (a_n-a_{n-1})^{-2}$, $W_{LNbnn+1max} = N|2se_{n-1}(z,q)se_n'(z,q)|^2(b_n-b_{n-1})^{-2}$. The electron transfer $W_e(E_m)$ was set to be 1. The Mathieu functions are from Ref. [13] at $q = 6$. The normalization factor is $N = 8.98\times10^{-2}$.